\documentclass[11pt]{article} 

\usepackage[utf8]{inputenc} 

\usepackage{geometry} 
\usepackage{graphicx} 

\usepackage[parfill]{parskip} 

\usepackage{cite}
\usepackage{booktabs} 
\usepackage{array}    
\usepackage{paralist} 
\usepackage{verbatim} 
\usepackage{subfig}   
\usepackage{tabularx}
\usepackage{footnote}
\usepackage{tablefootnote}

\usepackage{fancyhdr} 
\usepackage{sectsty}
\allsectionsfont{\sffamily\mdseries\upshape} 
\usepackage{times}
\usepackage{authblk}

\usepackage{graphicx}
\def\equationautorefname#1#2\null{Eq.#1(#2\null)}

\usepackage{amsmath}
\usepackage{amssymb}
\usepackage{amsthm}
\usepackage{hyperref}
\newcommand{\ii}{\mathrm{i}}

\newcommand{\xc}{\mathrm{xc}}

\newcommand{\rr}{\mathbf{r}}

\title{$GW$ vertex corrected calculations for molecular systems}
\author{Emanuele Maggio\thanks{emanuele.maggio@univie.ac.at}~}
\author{~Georg Kresse\thanks{georg.kresse@univie.ac.at}}
\affil{University of Vienna, Faculty of Physics and Center for
Computational Materials Science, Sensengasse 8/12, A-1090 Vienna, Austria}
\date{} 
\begin{document}
\maketitle

\begin{abstract}
	Hedin's scheme is solved with the inclusion of the vertex function ($GW\Gamma$) for a set of small molecules.
	The computational scheme allows for the consistent inclusion of the vertex both at the polarizability level and in the self-energy.
	A diagrammatic analysis shows that the self-energy formed with this four-point vertex does not lead to double counting of diagrams, that can be classified as direct "bubbles" and exchange diagrams.
	By removing the exchange diagrams from the self-energy, a simpler approximation is obtained, called $GW^{\rm{tc-tc}}$.
	Very good agreement with expensive wavefunction-based methods is obtained for both approximations.
\end{abstract}

\section{Introduction}
\label{sec:Intro}
	In the past decades many theoretical methods have been developed in the attempt to predict and rationalise molecular electronic structures.	
	Coupled cluster approaches are amongst the most widespread reference methods and are based on the exponential \textit{Ansatz} for the molecule's wavefunction \cite{Bishop1991,Bartlett2007}.
	The unfavourable scaling of these methods, however, makes them unsuitable for large scale calculations or predictive assessments for a large number of molecular systems.

\begin {figure} 
        	\centering
	        \includegraphics [width=12.5cm] {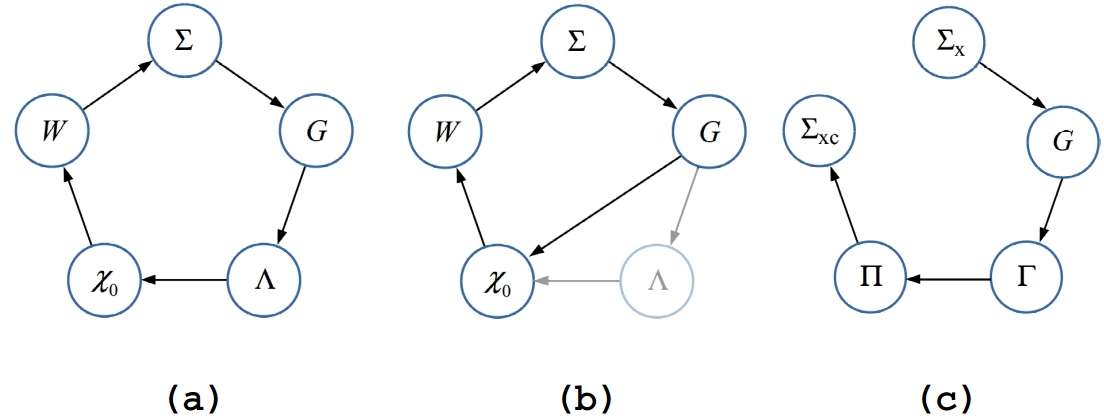}
        	\caption{Many-Body perturbation theory (MBPT) computational approaches. (a) Hedin's pentagon, (b) $GW$ approximation. (c) Computational scheme adopted in this work. See text for a definition of the symbols.}
	        \label{Fig:Hedin}
	\end{figure}

		In contrast to these well-established methods, many-body perturbation theory (MBPT) has emerged as a computational alternative for molecules \cite{Blase2011,Faber2011,Ke2011,Ren2012,Marom2012,Caruso2012a,Bruneval2013,Pham2013,Korbel2014,GW100}.
	The computational scheme first proposed by Hedin \cite{Hedin1965}, and summarised in \autoref{Fig:Hedin} (a-b), involves several quantities: starting with an initial guess for the Green's function ($G$) one determines the system's irreducible polarizability ($\chi_0$) and the screened interaction $W$.
	These quantities alone fully specify the self-energy operator $\Sigma$ in the first iteration of the scheme, which then produces an updated Green's function, as shown in panel (b).
	For later iterations the inclusion of vertex effects is in principle required through the  vertex function $\Lambda$. 
	The main difficulty in including the vertex is that it involves additional diagrams at each step, since the vertex is formally generated by the functional derivative of the self-energy with respect to the Green's function.
	Exploiting exact functional relations \cite{Schindlmayr1998a} can simplify the calculations only to a certain extent, and no purely numerical scheme can be implemented for the original method of Hedin. 
	It is understandable then that the $GW$ approximation (GWA) completely neglects  the vertex, thus iteratively applying the scheme in \autoref{Fig:Hedin}-b.
	This approximation is justified when exchange and correlation effects are completely absent in the reference state and only for the first iteration of the scheme, \textit{i.e.} when $\Sigma_{\xc}=0$, as for the Hartree case \cite{Morris2007}. 
	However, it has been widely used also on Hartree-Fock (HF) or density functional theory (DFT) starting points \cite{Klimes2014b,Bruneval2013,Wang2003,Sommer2012,Miyake2013,Cazzaniga2012}.

		Actual GWA calculations vary significantly in details.  The most common approximation is to create the starting orbitals using
	standard density functional theory and perform a single shot $G_0W_0$ calculation. This works well for extended systems, 
	but for small molecules this approximation tends
	to yield underbound highest molecular orbitals (HOMO). One way to cure this problem is to perform the calculations
	self-consistently.
	An update of the quasi-particle (QP) energies alone produces what is now often called the ev-$G_nW_0$ or the ev-$G_nW_n$ schemes, if the updated eigenvalues are included only in the Green's function or also in the screened interaction, respectively.
	Fully self-consistent $GW$ calculations \cite{Caruso2012a,Caruso2013a,Kutepov2016} require the calculation of the interacting Green's function $G(\omega) = (\omega -T-V_{\rm{ext}}-\Sigma)^{-1}$, with $T$ the kinetic energy operator and $V_{\rm{ext}}$ the external potential generated by the atomic nuclei and
	$\Sigma$ being the self-energy.
	The resulting quasi-particle equations involve an energy dependent non-Hermitian self-energy.
	To make this problem more manageable, quasi-particle self-consistent $GW$ approaches \cite{Faleev2004,VanSchilfgaarde2006} have been devised that disregard the non-Hermitian component of the self-energy and approximate its energy dependence \cite{Shishkin2007a}.
	Recent studies on selected organic molecules have shown that self-consistency significantly improves the quasi-particle energies\cite{Caruso2013a,Gallandi2016,Kaplan2016}.
	However, the HOMO is now on average too negative;  in other words self-consistent approaches tend to over-bind the electrons and yield a too large ionization potential (IP). As we will discuss below the inclusion of the vertex cures this shortcoming already in the first iteration of the scheme.

    Apart from the overestimation of the binding energies, self-consistency is also computationally fairly expensive, and other simple
    means to improve the QP energies have been considered as well. 
	One is the admixture of  exact exchange in the reference mean-field ground state calculation.   From a fundamental point of view,
	 this should certainly improve the occupied orbitals,  since for the standard local and semi-local density functionals that are commonly used as starting point 
	 the exchange-correction potential does not properly decay like one over the distance from the nuclei~\cite{Handy1969}. Also the HF one electron energies
	 are closer to the final $GW$ QP energies, so that an iterative solution of the $GW$ equations should not be necessary.
	 An elegant mean to achieve the correct $1/r$ decay of the nuclear-electron potential 
	 is using long-range  hybrid functionals \cite{Bruneval2013,Gallandi2016}, but standard hybrid functionals\cite{Kaplan2016}
	 as well as Hartree-Fock starting points have been used as well. 

	In the present work, we have also decided to use the HF reference point to initiate the $GW$ calculations. The reasons for this choice are summarized below.
	(i) From a practical point of view, this choice obviously allows for a rather straightforward comparison with wavefunction based methods 
	that usually rely on the HF reference point. Accordingly, we will use CCSD(T) reference data taken from the literature \cite{Bruneval2013}.
	(ii) There is evidence that HF theory is a reasonably accurate starting point for small molecules: 
	it is well known that, because of the Brillouin condition in HF theory, single excited configurations only contribute in second order in the perturbative expansion of the exact wavefunction, whereas for Br\"uckner orbitals their contribution is exactly zero to all orders.
	From a diagrammatic standpoint this translates into an identical cancellation of single excited determinants in lowest order for the ground state energy evaluated using either the HF or the Br\"uckner orbitals \cite{Kobe1971}.
	Since in localised systems the contribution of correlated higher order excitations is small \cite{Bhaskaran-Nair2016}, one can reasonably expect that the HF determinant has a large overlap with the exact many-body wavefunction, similarly to the Br\"uckner orbitals.
	(iii)  The main point of the present work is to investigate how important vertex corrections are. For the HF starting point, the vertex corrections
	can be evaluated exactly, as they are give by the functional derivative of the exchange  potential with respect to the Green's function. Hence we can unambiguously study the effect of vertex corrections.
	
	This brings us back to a point already alluded to above. $G_0W_0$@HF, as well as self-consistent $GW$ place the HOMO at too negative binding energies, or in other words, the IP is too large \cite{Bruneval2013}. The reason for this is fairly simple. In the GWA--- which implies that the random phase approximation is used
	to calculate the polarizability 
	---particle-hole ladder diagrams are not included. 
	The resulting $W$ underscreens the interaction between electrons because the energy to create an electron-hole pair is too large. 
	This largely explains why the IP is overestimated.
	The second issue affecting the GWA stems from the so-called self-screening error \cite{Nelson2007a,Fernandez2009a,Aryasetiawan2012}.
	In the $GW$ self-energy, one state can be occupied by two particles, since direct and exchange interactions are not treated on the same footing \cite{Romaniello2009}.
	This situation is reminiscent of the self-interaction problem that appears at the mean-field level in the Hartree theory. The inclusion of the exact exchange in the Hartree-Fock  theory completely cures this shortcoming on the mean-field level but neglects correlation. Including
	the functional derivative of the exchange with respect to the Green's function in Hedin's equations also eliminates the problem on the level of correlations. 
	
	To set the stage, we now summarize our computational approach. The initial step is a HF calculation for a set of molecules. 
	We then calculate the standard $G_0W_0$@HF QP energies for this set, and find as expected that this approximation overestimates the IP.
	We then present post-$GW$ calculations for the same set of molecules. 
	 In these calculations, the vertex corrections are included and given by the functional derivative of the exchange potential $\Sigma_\mathrm{x}$
	 with respect to the HF Green's function $G(\omega) =(\omega -T -V_{\rm{ext}}-V_H -\Sigma_\mathrm{x})^{-1}$
    (here  $V_H$ is the Hartree potential).
    The vertex is included either only in the polarizability ($G_0W_0^{\rm{tc-tc}}$)  or in the polarizability  and in the self-energy ($G_0W_0\Gamma$). In present work, only the diagonal components of the self-energy are calculated, and changes of the orbitals are concomitantly not considered,
    nor do we perform the calculations self-consistently, since this would in principle require a consistent update of the vertex as already emphasized above.	

		The inclusion of the vertex in the polarizability is equivalent to calculating the polarizabilty using time-dependent Hartree-Fock \cite{RevModPhys.36.844}, and the computational cost is also similar with both methods scaling like $O(N^6)$. 
	There are also formal similarities to the  Bethe-Salpeter equation (BSE) performed on top of a $GW$ reference state. The latter method  is commonly adopted for solid state systems \cite{Albrecht1998,Rohlfing2000}, but has also been used to study  the optical properties of  molecular systems \cite{Bruneval2015,Jacquemin2015}. 	
	To the best of our knowledge, vertex corrections have been routinely considered only for extended systems \cite{Bobbert1994,Delsole1994,Bruneval2005,Morris2007,Maebashi2011,Gruneis2014a,Kutepov2016}. For finite systems, calculations have been performed
	for atoms \cite{Shirley1993}, for very simple molecules (within the Tamm-Dancoff approximation) \cite{Kuwahara2016},
	or with the SOSEX \cite{Ren2015,Gallandi2016}, which approximates the vertex only in second order. 
	A local vertex $\Lambda_{\rm{LDA}}$ \cite{Hung2016} has also been used for a set of aromatic molecules, however, this simple two-point vertex behaves quite differently than the four-point many-body vertex used here.
	Among other things, the local approximation breaks fundamental self-energy symmetries and, albeit these effects are small for extended solids \cite{Delsole1994}, they might be more relevant for very inhomogeneous systems such as molecules.

		The work is structured in the following manner. In \autoref{sec:Theo} we derive simple equations for the vertex corrected $GW$, 
		reducing the set of equations to three equations that are reminiscent of the equation of motion. Contrary to Hedin's equation we use a four-point notation. 
	In \autoref{sec:2bleCounting} we report a diagrammatic analysis of the self-energy thus generated.
	The computational details are given in \autoref{sec:Comp}, and our results are finally presented and discussed in \autoref{sec:Res}, where we apply the $GW\Gamma$ method perturbatively to a set of molecules.

\section{Theory}
\label{sec:Theo}	
\subsection{Hedin's equations}
\label{sec:Hedin}	

		In principle, there are at least two complementary routes to improve on the MBPT approaches: either a self-consistent evaluation of the standard GWA or the inclusion of sub-leading diagrammatic contributions.
	Practical implementations have been directed primarely towards self-consistency (see \autoref{sec:Intro} for a discussion), whereas vertex corrections remain largely unaddressed, because of the high implementational as well as computational complexity. 

	To present our computational approach we will rely on a four point notation, for instance used  by Starke and Kresse \cite{Starke2012}, although in the present work we have decided to rearrange the indices to more easily connect with the available standard literature. 
	Also the specific succession of indices adopted here is easier to memorise. 
	Using a four index notation Hedin's equations can be written as
\begin{eqnarray}
 G(1,2) & =&  G_0(1,2) + G_0(1,3) \Sigma_{\xc}(3,4) G(4,2) \label{equ:h1} \\
 \Sigma_{\xc}(1,2) &= & \ii \,G(5,6) \Gamma(1,5,3,4) W(3,4,2,6) \label{equ:h2} \\
  \Pi_0(1,2,3,4) & = &- G(1,5) G(6,2) \Gamma(5,6,3,4) \label{equ:h3} \\ 
  W(1,2,3,4) & =& V(1,2,3,4) + \label{equ:h4} \\
           & + &V(1,2,5,6) \Pi_0(5,6,7,8) W(7,8,3,4) \nonumber \\
  \Gamma(1,2,3,4)& = &\delta(1,3)\delta(2,4) + \label{equ:h5} \\
     &-&  \ii \, I(1,2,5,6) G(5,7) G(8,6) \Gamma(7,8,3,4)  \nonumber
\end{eqnarray}
	In these equations, integrals over repeated indices are assumed. 
	We note that  if two four-point quantities are multiplied, for instance, $\Pi_0 W$ in Eq. (\ref{equ:h4}), then the integral is always over the two intermediate indices, in the equation below 5 and 6, and the order is easy to memorise:
\begin{equation}
\label{equ:matrixmul}
A(1,2,3,4) = \int \Pi_0(1,2,5,6) W(5,6,3,4) \, d5 \, d6. 
\end{equation}
	The advantage of using a four point notation is the following. 
	In Hedin's equations a numeral, for instance $1$, corresponds necessarily to a space-time and spin point $1=({\bf r}_1, t_1, \sigma_1)$;
	Hedin's original equations do not apply in orbital space or momentum space. 
	In the four point notation above, however, one can perform a unitary transformation of the position coordinate to reciprocal space or to any set of orthogonal orbitals without changing the equations.
	The slight disadvantage of the present ordering of the indices is that the first index and the fourth index in each four point object transform like covariant coordinates (transformed by say $U$), whereas the second and third indices transform like contra-variant coordinates (transformed by $U^\dagger$). 

		Before continuing we briefly reiterate the meaning of the individual objects.
	The four-point Coulomb interaction in space-time coordinates is $V(1,2,3,4)= \delta(1,2) \delta(3,4) \delta(t_1, t_4) v(\rr_1,\rr_4)$ generalising the usual two-point counterpart $v(\rr_1,\rr_4)=\frac{1}{|\rr_1-\rr_4|}$.
	$G_0(1,2)$ is the Hartree Green's function, generated by the corresponding Hartree self-energy: $\Sigma_0(1,2)=$ $V_H(1,2)=$ $\ii \delta(1,2) \delta(t_1,t_3) [v(\rr_1,\rr_3) G(3,3^+)]$ in space-time coordinates, $\Sigma_{\xc}(1,2)$ is the self-energy	including exchange as well as correlation terms. 
	$\Pi_0$ is the irreducible polarization (propagator); it can not be divided into two individual polarization propagators by cutting a single Coulomb line $V$. 
	$W$ is the screened interaction summing the Coulomb interactions to infinite order.
	Finally $\Gamma$ is the four-point vertex, which is completely specified by the kernel $I$: 
	\begin{equation}
	\label{eq:kern}
	I(1,2,3,4) = \frac{\delta \, \Sigma_{\xc}(1,2) }{\delta\,  G(3,4)}.
	\end{equation}

	There is an alternative way to write these equations that avoids the vertex altogether.
	We first introduce an auxiliary independent two-particle propagator $L_0(1,2,3,4)= -G(1,3)G(4,2)$.
	This allows to rewrite Eqs. (\ref{equ:h3}) and (\ref{equ:h5}) as
\begin{eqnarray}
\Pi_0(1,2,3,4) & = & L_0(1,2,5,6) \Gamma(5,6,3,4) \nonumber \\
\ \Gamma(1,2,3,4)& = &\delta(1,3)\delta(2,4) + \nonumber \\
     &+&  \ii \, I(1,2,5,6) L_0(5,6,7,8) \Gamma(7,8,3,4).  \nonumber
\end{eqnarray}
	We can now combine both equations  to a single equation 
\begin{equation}
\label{equ:pi0}
\Pi_0 = L_0 + L_0  (\ii \, I) \Pi_0 \quad\leftrightarrow   \quad\Pi_0^{-1} = L_0^{-1} - \ii \, I.
\end{equation}
	Here we have suppressed the indices and the matrix multiplications need to be done according to the rule stated above (Eq. \ref{equ:matrixmul}). 
	For the inversion, it is understood that the first two and last two indices of each four point quantity are combined to a super index and the matrices are inverted using the two super indices as row and column indices.
	Likewise, we can introduce the full polarization propagator
\begin{equation}
\label{equ:pi}
\Pi = \Pi_0 +  \Pi_0 V \Pi  \quad\leftrightarrow   \quad\Pi^{-1} = \Pi_0^{-1} - V.
\end{equation}
	It is a simple matter to see that these two equations can be further combined into a single equation, reading 
\begin{equation}
\label{equ:pifinal}
\Pi = L_0 +  L_0 (V+ \ii \, I) \Pi \leftrightarrow\Pi^{-1} = L_0^{-1} - \ii \, I- V.
\end{equation}
	The proof is most easily done inspecting the right hand inverted Dyson like equations in the previous three equations.
	Eq. (\ref{equ:pifinal}) is precisely the polarization propagator as calculated by means of the ``Bethe-Salpeter equation'' (BSE) in most solid state codes.
	In fact, in the preceding lines, we have backtracked the calculations performed by Starke and Kresse, where Hedin's equations were derived from the equation of motion and the BSE.

		We can now proceed with $W$ and the self-energy.
	By expanding  the right hand side of Eq. (\ref{equ:h4}) to infinite order in $V$ one can identify $\Pi$ and rewrite Eq. (\ref{equ:h4}) as
\begin{equation}
\label{eq:W}
 W = V + V \Pi V. 
\end{equation}
	The final issue is to rewrite Eq. (\ref{equ:h2}), specifically the four point term $\Gamma W = \Gamma(1,2,3,4) W(3,4,5,6) $ such that it involves only the already calculated polarization propagators.
	Using Eq. (\ref{eq:W}) for $W$ and inserting the vertex from Eq. (\ref{equ:h5}), we obtain:
\begin{equation}
\begin{split}
 \Gamma W  & = ( 1 - (\ii I) L_0 -(\ii I) L_0 (\ii I) L_0 + ...) (V + V \Pi V )  \nonumber  \\
  & = ( 1 + (\ii I) \Pi_0) (V + V \Pi V )  \\
  & =  V + V \Pi V +  (\ii I) (\Pi_0 +\Pi_0 V  \Pi ) V  \\
  & =  V + V \Pi V +  (\ii I) \Pi  V =  V + (V +\ii I) \Pi V.
\end{split}
\end{equation}
	From the first to the second line we have used Eq. (\ref{equ:pi0}), and from the third to the fourth  line we have used Eq. (\ref{equ:pi}).

		In summary, we can rewrite Hedin's equation in the much more compact form
\begin{eqnarray}
\Pi & = &  L_0 +  L_0 (V+ \ii \, I) \Pi  \label{eq:PiBSE} \\
\Sigma_{\xc} & =& \ii \, G \Big( V + (V +\ii I) \Pi V \Big ) \label{equ:GWGamma} \\
 G & =&  G_0 + G_0 \Sigma_{\xc} G, \label{equ:greens}
\end{eqnarray}
	where the contraction of the two-point Green's function and a four-point quantity is here defined as
\begin{equation}
 \Sigma_\mathrm{x}(1,3) =\ii (GV)(1,3) =  \int G(2,4) V (1,2,3,4) d 2 \, d 4.
\end{equation}
	In these equations the somewhat arbitrary distinction between the Coulomb kernel $V$ and the remaining interaction kernel $\ii I$ has been  dropped.
		These equations are essentially equivalent to the compact equations given by Starke and Kresse but have been derived here  from Hedin's equations instead of the more fundamental equation of motion.
	We note that the same equation for the self-energy can be also found in Ref.  \cite{Romaniello2012} (albeit without derivation) and is seemingly common knowledge in quantum field theoretical publications \cite{Held2011,Ayral2013}.

		A few comments are in place. As already emphasized, $\Pi$ in Eq. (\ref{eq:PiBSE}) is the polarization propagator (or polarizability) that many solid state BSE codes calculate. It describes how test charges are screened by the electronic system, hence it is often referred to as 
test-charge test-charge (tc-tc) polarizability. The self-energy Eq. (\ref{equ:GWGamma}) describes
the effects of the many-body system on an added or removed electron, and Eq. (\ref{equ:greens}) is the related Green's function.
	The first term on the right hand side of Eq. (\ref{equ:GWGamma}) is just the exchange potential ($GV$), whereas the second term describes, in principle,  all correlation effects. If $iI$ is neglected in
the self-energy  (\ref{equ:GWGamma}) but included in (\ref{eq:PiBSE}) the approximation is commonly
referred to as $GW^{\rm{tc-tc}}$. 

 Although this set of equations is in principle exact, approximations need to be made for the interaction kernel $I$ specified in  \autoref{eq:kern}.
	In our case, we perform calculations perturbatively on the Hartree-Fock reference state, \textit{i.e.} we start the Hedin scheme with the sum of the Hartree ($\Sigma_0$) and exchange self-energy ($\Sigma_\mathrm{x}$): $\Sigma= \Sigma_0 + \Sigma_\mathrm{x}$.  
        Then the interaction kernel is simply the functional derivative of the the exchange self-energy $\Sigma_\mathrm{x}$ and thus explicitly given by
\[
 \ii I(1,2,3,4) = - \delta(2,4) \delta(1,3) \delta(t_1, t_2) v(\rr_1,\rr_2),
 \]
	where $v( \rr_1,\rr_2)$ is the bare Coulomb interaction between particles. The polarization propagator is then just the  polarizability of time-dependent Hartree-Fock  \cite{RevModPhys.36.844}.
	The approximation employed in this study is also related to the random phase approximation with exchange (RPAx) first proposed by Szabo and Ostlund \cite{Szabo1977}.
	There has been a recent revival in interest for the RPAx \cite{Hesselmann2010,Bleiziffer2012,Furche2013,Nguyen2014,Mussard2015,Colonna2016}, however, previous studies have focused on the evaluation of correlation energies, rather than the self-energy considered here.

	       Finally, we comment on the use of the $GW$ reference state instead of the Hartree-Fock starting point employed here. For the $GW$ reference state, the interaction kernel is given by the functional derivative of the $GW$ self-energy with respect to the Green's function. This makes the interaction kernel more complicated. Specifically, (i) $W$ is frequency dependent, and (ii) since $W$ itself depends on the Green's function, the derivatives of $W$ with respect to $G$ should be accounted for. The resulting diagrams are sometimes refereed to as ``butterfly'' diagrams.
 A common approximation is to neglect the dependence of $W$ on the Green's functions, and to approximate the frequency dependence of $W$ by an instantaneous interactions. The most common approximation for the interaction kernel is 
\begin{equation}
\label{eq:BSEkern}
 \ii I(1,2,3,4) = - \delta(2,4) \delta(1,3) \delta(t_1, t_2) w(\rr_1,\rr_2),
 \end{equation}
	with $w(\rr_1,\rr_2)$ being the random phase approximation (RPA) screened Coulomb interaction at $\omega=0$.
 These approximations are also commonly employed in solid state BSE codes \cite{Albrecht1998,Rohlfing2000,Fuchs2008,Sander2015,Ljungberg2015}, and our present $GW\Gamma$ code can be adopted to this case. We will report on such calculations in forthcoming publications.

\subsection{Bethe-Salpeter equation}
\label{sec:BSE}	

	As already emphasized above, we treat all the interactions as instantaneous in the present case. For the Hartree-Fock reference point this is exact, but
    it constitutes an approximation, if one were to start from the self-energy of the $GW$ approximation.
	In full generality, the BSE can be Fourier transformed into the frequency domain and its kernel becomes a function of three independent frequency variables $I=I(\omega,\omega',\tilde{\omega})$, with the screened interaction carrying a frequency dependence of $\omega'-\tilde{\omega}$ \cite{Romaniello2009a}.
	Using an instantaneous interaction obviously yields a kernel $I$ that does not depend on the frequency,
        since the Fourier transformation of a $\delta$-function in time is a constant in frequency space.
        More detailed discussions can be found in literature \cite{Hanke1979,Hanke1980,Strinati1988}. 
	Then, in an orbital representation the full kernel $V +\ii I$ is given by the matrix elements of the matrices $\mathbf{A'}$ and $\mathbf{B}$:
	        \begin{align}
                        A'_{iajb}=& \langle aj|V|ib \rangle - \langle aj|V|bi \rangle, \label{eq:RR}\\
                        B_{iajb}=& \langle ab|V|ij \rangle - \langle ab|V|ji \rangle.\label{eq:RA}
                \end{align}
	The BSE reduces  to a generalised eigenvalue problem \cite{Fuchs2008},	where the eigenvalues $\Omega_{\lambda}$ correspond to the optical transition energies, and the matrix $\mathbf{A}$ is formed from $\mathbf{A'}$ adding the independent particle energy differences $\Delta E_{ia}=\epsilon_a - \epsilon_i$ to the diagonal elements:
		\begin{align}
                \label{eq:BSE}
                \begin{pmatrix}
                         \mathbf{A}   & \mathbf{B}\\
                         \mathbf{B}^* & \mathbf{A}^*
                \end{pmatrix}
		\left |
                \begin{matrix}
                         \mathbf{X}_\lambda\\
                         \mathbf{Y}_\lambda 
                \end{matrix}
		\right \rangle =
                         \Omega_{\lambda}
		\left |
                \begin{matrix}
                         \mathbf{X}_\lambda\\
                         \mathbf{Y}_\lambda
                \end{matrix} 
		\right \rangle. 
                \end{align}
  		At the $\Gamma-$point the matrix elements of $\mathbf{A}$ and $\mathbf{B}$ can be chosen to be real-valued and since the matrix $\mathbf{A}$ is Hermitian and $\mathbf{B}$ is symmetric \cite{Onida2002}, the generalised eigenvalue problem (EVP) above can be recast as an Hermitian EVP, for which standard solvers are available.
		A concise expression for the resulting polarisation propagator's spectral representation reads \cite{Ring1980,Maggio2016}:
	\begin{align}
        \label{eq:PiSR1}
        \mathbf{\Pi} (\omega) = \mathbf{Z} \left ( \omega - \mathbf{\Omega} \right )^{-1} \mathbf{\Delta}\mathbf{Z}^\dagger, 
        \end{align}
	with the following definitions for the matrices: 
	\begin{align}
                \mathbf{Z}=
                \begin{pmatrix}
                \mathbf{X} & \mathbf{Y}^* \\
                \mathbf{Y} & \mathbf{X}^*
                \end{pmatrix},
		\mathbf{\Delta}=
                \begin{pmatrix}
                \mathbf{1} & \mathbf{0} \\
                \mathbf{0} & \mathbf{-1}
                \end{pmatrix},
                \mathbf{\Omega}=
                \begin{pmatrix}
                \mathrm{diag}\{\Omega_\lambda \} & \mathbf{0} \\
                \mathbf{0} & \mathrm{diag}\{-\Omega_\lambda \}
                \end{pmatrix}.
        \end{align} 
	The matrix $\mathbf{Z}$ contains all the individual eigenvectors $\mathbf{X}_{\lambda}$ and $\mathbf{Y}_\lambda$, similarly the matrix $\mathbf{\Omega}$ includes resonant and antiresonant transition energies; the transition energies for the resonant and antiresonant branch are identical (apart from a sign) as a consequence of the downfolding of the original BSE into an Hermitian EVP.
	For definitiveness we choose $\Omega_\lambda >0$.
	Since the matrix $\omega - \mathbf{\Omega}$ is diagonal in the orbital representation, it can be commuted across; the resulting product $\mathbf{Z} \mathbf{\Delta} \mathbf{Z}^\dagger$ can be decomposed into the difference of two matrices:
	\begin{align*}
                \mathbf{Z \Delta Z}^\dagger =
                \begin{pmatrix}
                \mathbf{XX}^* & \mathbf{XY}^* \\
                \mathbf{YX}^* & \mathbf{YY}^*
                \end{pmatrix} -
                \begin{pmatrix}
                \mathbf{Y^*Y} & \mathbf{Y^*X} \\
                \mathbf{X^*Y} & \mathbf{X^*X}
                \end{pmatrix}. 
        \end{align*} 
	The resonant and antiresonant contributions can be more easily singled out by re-writing the matrices above as an external product of the eigenstates of \autoref{eq:BSE}.
	The polarisation propagator then reads:
	\begin{align}
	\label{eq:PiSR2}
	\Pi(\omega) = \sum_{\lambda} \frac{ \left |\begin{matrix} \mathbf{X}_\lambda \\ \mathbf{Y}_\lambda \end{matrix} \right \rangle \langle \mathbf{X}_\lambda^* \mathbf{Y}_\lambda^*| }{\omega - \Omega_\lambda} - 
	\frac{\left |\begin{matrix} \mathbf{Y}_\lambda^* \\ \mathbf{X}_\lambda^* \end{matrix} \right \rangle \langle \mathbf{Y}_\lambda \mathbf{X}_\lambda| }{\omega + \Omega_\lambda}.
	\end{align}
	This expression can now be inserted into \autoref{equ:GWGamma} to give the greater and lesser components of the correlation self-energy, corresponding to the propagation of a particle ($p$) and a hole ($h$), respectively: 
	\begin{align}
	\label{eq:SigmaR}
	\langle n | \Sigma_p (\omega) | n \rangle & =\sum_\lambda \langle n |\mathbf{V} |\mathbf{X}_\lambda + \mathbf{Y}_\lambda \rangle \langle \mathbf{AX}_\lambda + \mathbf{BY}_\lambda|  n \rangle \left ( f(\epsilon_n) -1 \right ) \delta \left ( \epsilon_n + \Omega_\lambda - \omega \right ) \\
	\label{eq:SigmaA}
	\langle n | \Sigma_h (\omega)| n \rangle & =\sum_\lambda \langle n |\mathbf{V} |\mathbf{X}_\lambda + \mathbf{Y}_\lambda \rangle \langle \mathbf{AY}_\lambda + \mathbf{BX}_\lambda|  n \rangle f(\epsilon_n) \delta \left ( \epsilon_n - \Omega_\lambda - \omega \right ). 
	\end{align}
	$f(\epsilon_n)$ in the previous equations is the occupancy for the energy level $\epsilon_n$ and the summation goes over the set of particle-hole excitations $\lambda$.
	The composite index $\lambda$ is constructed by considering all possible combinations of the single particle indices $(i,a)$, with $i$ and $a$ belonging to the occupied and unoccupied orbital manifold; in the following $n$ represents a generic orbital.
	The total correlation part of the self-energy is then obtained by taking the Hilbert transform of the two components:
	\begin{align}
	\label{eq:Sigmac}
	\langle n | \Sigma_c (\omega) | n \rangle & = \int \frac{d\omega'}{2} \frac{\langle n | \Sigma_p (\omega') - \Sigma_h (\omega')| n\rangle }{\omega - \omega' +\ii \eta ~\mathrm{sgn}(\omega - \mu)}.
	\end{align}
	Expressions in Eqs. (\ref{eq:SigmaR}) and (\ref{eq:SigmaA}) reduce to the usual RPA limit \cite{Tiago2006}, if the exchange term in $\mathbf{A'}$ and $\mathbf{B}$ is omitted.

\subsection{Diagrammatic analysis}
\label{sec:2bleCounting}	
\begin {figure} 
        	\centering
	        \includegraphics [width=7.5cm] {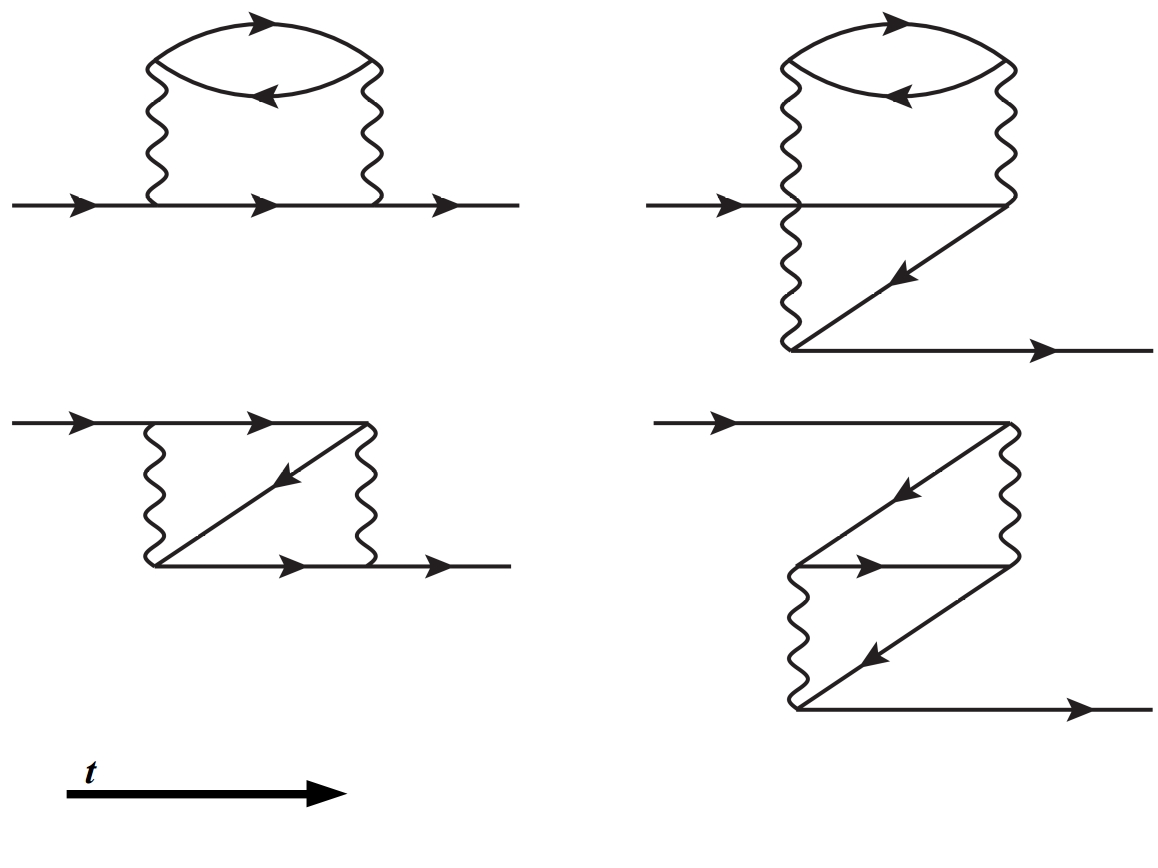}
        	\caption{Self-energy diagrams for the lowest order approximation; these are obtained by replacing $\Pi$ with $L_0$ in \autoref{equ:GWGamma}. Wavy lines represent the bare Coulomb interaction. The direction of time is fixed by the thick arrow and it is understood in the following diagrams.}
	        \label{Fig:SE_2ndO}
	\end{figure}

		We start our diagrammatic analysis with the lowest order expressions: these can be obtained by replacing the fully interacting polarisation propagator $\Pi$ in \autoref{eq:PiBSE} with the non-interacting counterpart $L_0$.
	The corresponding diagrams are shown in \autoref{Fig:SE_2ndO}.
	In the top row of \autoref{Fig:SE_2ndO} we are reporting the correlation contributions in \autoref{equ:GWGamma} that stem from the term labelled with $V$, in the bottom row the terms generated by the kernel $\ii I$ are shown.
	The two columns portray the different orientations of the intermediate state $n$ with respect to the incoming state $n'$ (which in the diagram is represented by the external legs which have not been amputated for clarity): in the left column $n$ and $n'$ belong to the same manifold (they are both either occupied or unoccupied), in the right column $n$ and $n'$ belong to different manifolds.
	It is obvious, by inspecting the diagrams, that the two contributions (in the top and bottom row, called respectively bubble and exchange in the following) are non equivalent in lowest order and that by cutting a particle-hole pair the diagram does not break up into disjoint pieces.
	This lower order contribution results in the MP2 self-energy, evaluated for instance in Ref. \cite{Gruneis2014a} for extended systems.

\begin {figure} 
        	\centering
	        \includegraphics [width=7.5cm] {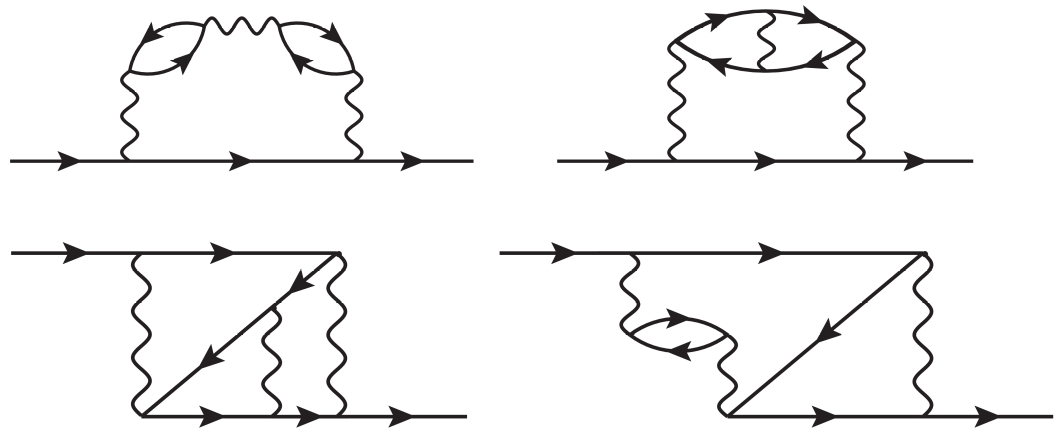}
        	\caption{Self-energy diagrams for the third order contributions, only diagrams corresponding to states $n, n'$ belonging to the same manifold are shown. Wavy lines represent the bare Coulomb interaction.}
	        \label{Fig:SE_3rdO}
	\end{figure}

\begin {figure} 
        	\centering
	        \includegraphics [width=9.0cm] {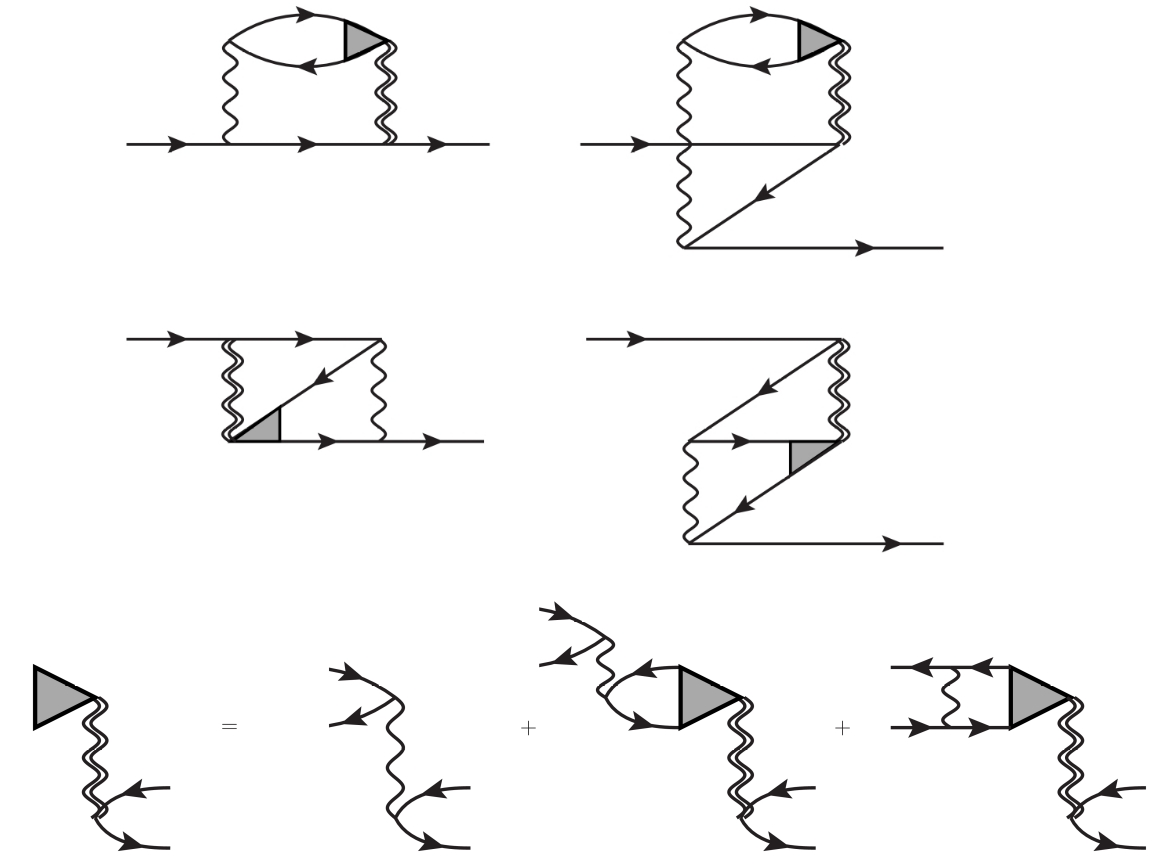} 
        	\caption{(Top) Self-energy diagrams for the infinite order resummation. A pictorial definition for the renormalised interaction lines (shaded triangles and double wavy lines) is given in the bottom line; these terms contain an infinite number of particle-hole interaction lines summed up by solving the BSE.}
	        \label{Fig:SE_infO}
	\end{figure}

		If we now switch on the interaction between particles and holes in the polarisation propagator we obtain additional classes of diagrams, as shown in \autoref{Fig:SE_3rdO} in third order in the interaction, \textit{i.e.} setting $\Pi=L_0 (V+\ii I) L_0$.
	Two of the third order diagrams belong to the same class as their second order counterparts, these are shown in the left column of \autoref{Fig:SE_3rdO}.
	Additionally, the "mixed contributions" shown in the right column represent the insertion of an exchange interaction in the bubble term (top) and the insertion of a bubble in the exchange diagram (bottom).
	The polarisation propagator, however, is obtained by a resummation to infinite order of the corresponding diagrams \textit{via} the solution of the BSE.
	These are represented in \autoref{Fig:SE_infO}, where we are showing both orientations for the intermediate particle for completeness. 
	Starting from the top left diagram in \autoref{Fig:SE_3rdO} it is easy to identify an infinite resummation of this class with the usual RPA approximation, where the so-called bubble diagrams are summed up; this translates into a screened interaction (denoted by a double wavy line) showing  up in the corresponding diagram in \autoref{Fig:SE_infO}.
	Likewise, an infinite number of bubble insertions in the exchange diagram of \autoref{Fig:SE_3rdO} (bottom right) returns an additional screened line.
	The infinite resummation of the exchange contributions (top right and bottom left in \autoref{Fig:SE_3rdO}), on the other hand, takes into account the interaction between particles and holes in the polarisation propagator and this is represented in \autoref{Fig:SE_infO} by a shaded triangle.
	We should now point out that in the final self-energy diagrams, the resummation to infinite order of the polarisation propagator generates a frequency dependent, dynamical screened interaction vertex (pictorially defined at the bottom of \autoref{Fig:SE_infO}).
	Considering the resulting diagrams it is easy to see that they result in proper, irreducible self-energy insertions: the presence of bubble and exchange contributions in the kernel assures that two topologically distinct classes are obtained (shown in the top and middle row of \autoref{Fig:SE_infO}) and that for either class there is no external polarisation insertion.

	As it was mentioned in the introduction, the iteration to self-consistency modifies the diagrammatic structure presented up to this point (see \autoref{Fig:Hedin}-a). This is one reason why we do not consider self-consistency  in the present work. As explained  at the end of Sec. \ref{sec:Hedin}, by adopting some further
        approximations, it is however possible to maintain the simple algebraic structure and extend the present 
        approach to $GW\Gamma$
        calculations on top of a $GW$ reference state.
		
		Finally, we mention that the exclusion of the exchange diagrams in the self-energy can be obtained by formally replacing the electron in the intermediate state $n$ with a classical test-charge. 
	In what follows we will refer to this level of theory as to the $GW^{\rm{tc-tc}}$ approximation.
	This approximation neglects the $\ii I$ term in \autoref{equ:GWGamma}, but includes it in \autoref{eq:PiBSE}.

\section{Computational procedure}  
\label{sec:Comp}
\begin{figure} 
        	\centering
	        \includegraphics [width=7.5cm] {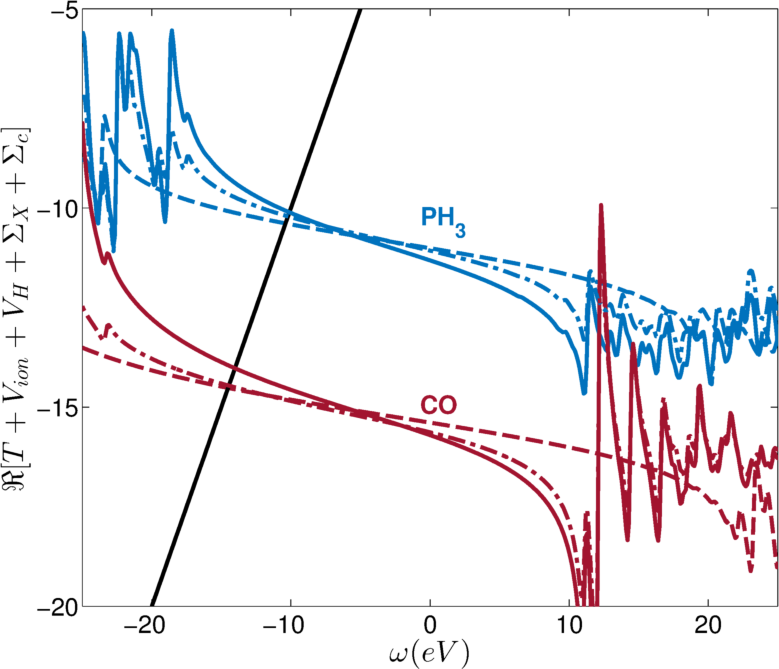}
        	\caption{Self-energy for the HOMO level of $\rm{PH_3} $ (blue lines) and CO (red lines) evaluated at the $G_0W_0$ (dashed lines), $G_0W_0^{\rm{tc-tc}} $ (solid lines) and $G_0W_0\Gamma$ (dashed-dotted lines) level. The solid black line is the bisector line of equation $y=\omega$. }
	        \label{Fig:SE_PH3_CO}
	\end{figure}

    		Calculations were performed on a subset of molecules studied in Ref. \cite{Bruneval2013}. For consistency we adopted the same molecular geometries as obtained in Ref.\cite{Curtiss1998} at the MP2/6-31G(d) level of theory.
        The initial step was always a self-consistent Hartree-Fock calculation. Since the solution of the Bethe-Salpeter equation is very demanding if many virtual orbitals are included, we used natural orbitals to represent the unoccupied states.
        In our procedure we roughly followed the work of Gr\"uneis \textit{et al.} \cite{Gruneis2014a}, but instead of using the MP2 density matrix, we calculated the RPA density matrix using the cubically scaling $GW$ code \cite{Liu2016}. 
        We then diagonalized the virtual-virtual sub-block and sorted the natural orbitals by their occupancy.
        In this step, the number of orbitals is reduced by a factor four compared to the plane wave basis set size specified by the cutoff $E_c$ and reported in \autoref{tab:G0W0}.
        We note that the occupied ($N_{\rm{occ}}$) as well as few unoccupied states are not modified but kept fixed at the level of the HF canonical orbitals.
	The number of unoccupied orbitals that are kept fixed at their Hartree-Fock level is set to 0.8$\times N_{\rm{occ}}$.
        By comparing calculations using all canonical orbitals and natural orbitals for selected systems, we found that the reduction of the unoccupied states introduces errors that are smaller than about 10~meV. 
        After determining the natural orbitals, the Hartree-Fock Hamiltonian is again diagonalized using the natural orbitals as basis \footnote{in VASP this is performed by selecting \texttt{ALGO=SUBROT}}. 
	Using this basis, the Bethe-Salpeter equation is solved for the polarization propagator, and the diagonal elements of the self-energy are calculated in the orbital basis.
	The self-energy is evaluated on the real frequency axis sampled with a uniform grid of 0.1 eV spacing. 
	The complex broadening is set to twice the frequency spacing and the resulting self-energy $\langle n |\Sigma_c(\omega) |n \rangle$ (shifted by the kinetic, ionic, Hartree and Fock exchange energy contributions) is piecewise linearly interpolated in each interval to determine the quasi-particle energies as the intersection with the bisector line as shown in \autoref{Fig:SE_PH3_CO}.
	The quasi-particle shifts ($\delta(\varepsilon^{\rm{QP}})$ in the following) are then obtained by subtracting the corresponding mean-field single particle energies.
    	To minimise the impact of image charges due to the finite simulation box, the Hartree-Fock calculations were repeated increasing the simulation cell up to 25 {\AA}  in each linear dimension and then correcting for the residual local potential present at the edge of the cell.
	The quasi-particle calculations were performed in a smaller simulation box (8 {\AA} in size).
	The volume dependence of the quasi-particle energies has been carefully analysed in our previous publication \cite{GW100vasp} and it was found to be neglible for the systems considered here.
	The quality of this approximation can be assessed for the case of $G_0W_0$ for the HF reference state by comparison with the localised basis set calculations in Ref. \cite{Bruneval2013} and reported in \autoref{tab:G0W0}.

    	The final quasi-particle energies were obtained by adding to the cell size converged (HF) single particle energies the quasi-particle shifts and the basis set corrections, according to the expression:
    	\begin{align} 
	\label{eq:eQP}
    	\left . \varepsilon_n^{\rm{QP}} \right |_\infty = \left . \varepsilon_n^{\rm{HF}}\right |_N + \delta(\varepsilon_n^{\rm{QP}} ) + C_N(\varepsilon_n^{\rm{QP}}).
    	\end{align}
	The basis set correction $C_N$ is defined as: $C_N(\varepsilon)=\left . \varepsilon \right |_\infty- \left . \varepsilon \right |_N$, where the subscripts indicate the number of natural orbitals ($N$) or the extrapolated value ($\infty$). 
	In the next section it will be shown how the interplay of quasi-particle energy shift and the basis set correction affects the final quasi-particle levels.  

    	One of the advantages of plane wave basis sets is the full control of the basis set completeness  by specifying their kinetic energy cutoff $E_c$. \footnote{in VASP this specified by the flag \texttt{ENCUT}}
	To estimate the basis set convergence, we increased $E_c$ by a factor 1.3 and 1.5 beyond the value specified in \autoref{tab:G0W0} and extrapolated assuming that the QP shifts converge like one over the basis set size \cite{Klimes2014b}.

\section{Results and discussion}
\label{sec:Res}

		In this section, we apply the computational scheme presented above to a subset of the molecules considered in Refs. \cite{Bruneval2013,Curtiss1998}.
	As a first check on the accuracy of the computational procedure we compare in \autoref{tab:G0W0} the HOMO quasi-particle energies at the $G_0W_0$ level of theory with those obtained by Bruneval and Marques with Gaussian type orbitals (GTOs) using cc-pVQZ basis sets \cite{Bruneval2013}.

	We immediately point out that we expect differences  of the order of 100 meV between PW and GTO $GW$ results.  First, pseudopotentials can introduce errors and second the GTO results were not extrapolated to the complete basis set limit.
	Concerning the pseudpotential error, we know from previous work that flourine containing compounds are affected most strongly with errors of about  $\approx$100~meV, owing to the strong degree of localisation for 2$p$ electrons. With the  non-normconserving PAW potential used here, we somewhat underestimate the degree of localization for flourine, as explained elsewhere \cite{GW100vasp}. Concerning the basis set extrapolation of the GTO results, 
	its convergence was studied in Ref. \cite{Bruneval2013} up to the "correlation-consistent" pV5Z basis set for the specific case of CO; the HOMO energy difference between this basis set and the quadruple zeta employed for the calculations of all the remaining molecules is about $\sim$70 meV. 
	Since the error decays like $1/C_n^3$, where $C_n$ is the cardinal number (\textit{i.e.} four for quadruple basis sets), the quadruple zeta basis set results reported in  Ref. \cite{Bruneval2013} are expected to have errors of the order of $2 \times 70~$meV$\approx 140~$meV.
	The difference between our plane-wave (PW) results and the GTO ones is typically comparable with this order of magnitude, with the exception of two outliers: CS and P$_{\text{2}}$.
	For CS we observe that in going from the HF to the $G_0W_0$ level there is a rearrangement of the orbital energies, with the HOMO-1 and the HOMO exchanging their positions.
	We can therefore speculate that the value reported in the literature refers to a lower lying quasi-particle level that corresponds to the HOMO orbital at the mean-field level.
	For this reason we have included also this quasi-particle energy in \autoref{tab:G0W0} for comparison.
	On the other hand, for P$_{\text{2}}$, we notice how the finite box effects are particularly strong in this case, with a downward shift exceeding 1 eV for the HF HOMO level  as the cell edge is increased from 8 to 25 \AA.
	It is, however difficult to draw a solid conclusion on the origin of the observed mismatch, also considering that in Ref. \cite{Bruneval2013} P$_{\text{2}}$ fails to follow any of the trends observed for the remaining molecules.    
	
\begin{table*}
\caption{ $G_0W_0$ quasi-particle energies for the highest occupied orbital (HOMO) evaluated on a Hartree-Fock reference state. 
In the second column the smallest plane-wave cutoff $E_c$ (and the corresponding number of plane waves $N_{\rm{pw}}$) used in the extrapolation is given. 
The last column reports the negative of the experimental ionisation potential (IP), vertical values are in italics. 
Plane wave extrapolated results include also finite box corrections as specified in \autoref{sec:Comp}.
All values in eV.}
\label{tab:G0W0}
\begin{tabular}{l c l c c c c c }
\hline \hline
      molecule & $E_c $ & ($N_{\rm{pw}}$)  & $G_0W_0  $             &   $G_0W_0$ \cite{Bruneval2013} & $\Delta$    & CCSD(T) \cite{Bruneval2013}  & Expt. \\ 
               &        &            &  PW                          & GTO                      &  PW $-$ GTO  & GTO                    &       \\   
               &        &            &                              & cc-pVQZ                  &              & cc-pVQZ                &       \\
\hline 
 H$_{\text{2}}$	 		&  300.0 & (8620)     & -16.72    &    ---       &   ---       &   -16.39 \cite{Koval2014}   &  -15.43       \cite{McCormack1989} 	 \\
 Li$_{\text{2}}$    		&  112.1 & (1940)     & -5.29     &   -5.36      &   0.07      &   -5.17                     &  -4.73        \cite{Dugourd1992}    	 \\
 N$_{\text{2}}$	    		&  420.9 & (10060)    & -16.56    &   -16.48     &  -0.08      &   -15.49                    &  -15.58       \cite{Trickl1989}   	 \\
 P$_{\text{2}}$	    		&  255.0 & (9180)     & -11.31    &   -10.57     &  -0.74      &   -10.76		     &  -10.62       \cite{Bulgin1976}         \\
 Cl$_{\text{2}}$    		&  262.5 & (9780)     & -12.08    &   -12.01     &  -0.07      &   -11.62                    &  {\em -11.49} \cite{Dyke1984}           \\
                                                                                                                                                                     
 HF		    		&  487.7 & (12540)    & -16.29    &   -16.39     &   0.10      &   -16.09                    &  {\em -16.12} \cite{Banna1975}          \\
 LiH	            		&  300.0 & (8620)     & -8.26	 &   -8.2       &  -0.06      &   -7.94                     &  -7.90        \cite{NIST2015}      	 \\
 NaCl		    		&  262.5 & (7080)     & -9.51	 &   -9.36      &  -0.15      &   -9.13                     &  {\em -9.80}  \cite{Potts1977}	  	 \\
 ClF				&  487.7 & (12540)    & -13.49    &   -13.32     &  -0.17      &   -12.82		     &  {\em -12.77} \cite{Dyke1984}           \\
                                                                                                                                                                     
 CH$_{\text{4}}$    		&  414.0 & (9800)     & -14.95    &   -14.86     &  -0.09      &   -14.4                     &  {\em -13.6}  \cite{Bieri1980}          \\
 C$_{\text{2}}$H$_{\text{4}}$	&  414.0 & (9800)     & -10.91    &   -10.85     &  -0.06      &   -10.69                    &  {\em -10.68} \cite{Bieri1980}          \\
 C$_{\text{2}}$H$_{\text{2}}$	&  414.0 & (9800)     & -11.73    &   -11.65     &  -0.08      &   -11.42                    &  {\em -11.49} \cite{Bieri1980}          \\
 CH$_{\text{3}}$Cl  		&  414.0 & (9800)     & -11.90    &   -11.74     &  -0.16      &   -11.41                    &  {\em -11.29} \cite{Kimura1981} 	 \\
 CH$_{\text{3}}$OH		&  434.4 & (10460)    & -11.71    &   -11.69     &  -0.02      &   -11.08		     &  {\em -10.96} \cite{Vorobev1989}        \\
 CH$_{\text{3}}$SH		&  414.0 & (9800)     & -9.93	 &   -9.81      &  -0.12      &   -9.49		     &  {\em -9.44}  \cite{Cradock1972}        \\
 SiH$_{\text{4}}$		&  300.0 & (8620)     & -13.40    &   -13.31     &  -0.09      &   -12.82		     &  {\em -12.3}  \cite{Roberge1978}        \\
                                                                                                                                                                     
 CO		    		&  434.4 & (10460)    & -15.03    &   -14.97     &  -0.06      &   -14.05                    &  {\em -14.01} \cite{Potts1974}          \\
 CO$_{\text{2}}$    		&  434.4 & (10460)    & -14.35    &   -14.38     &   0.03      &   -13.78                    &  {\em -13.77} \cite{Eland1977}          \\
 SiO				&  434.4 & (10460)    & -12.05    &   -11.98     &  -0.07      &   -11.55		     &  -11.3        \cite{Nakasgawa1981}      \\
 SO$_{\text{2}}$		&  434.4 & (20380)    & -13.20    &   -13.12     &  -0.08      &   -12.41		     &  {\em -12.50} \cite{Kimura1981}         \\
 CS  				&  414.0 & (9800)     & -12.63    &   -13.08     &   0.45      &   -11.45		     &  -11.33       \cite{King1972}           \\
 CS (*)				&        &            & -13.19    &   -13.08     &  -0.11      &   -11.45		     &  -11.33       \cite{King1972}           \\
 H$_{\text{2}}$O    		&  434.4 & (10460)    & -13.10    &   -13.04     &  -0.06      &   -12.64                    &  {\em -12.62} \cite{Kimura1981}         \\
 H$_{\text{2}}$O$_{\text{2}}$	&  434.4 & (10460)    & -12.12    &   -12.13     &   0.01      &   -11.49                    &  {\em -11.70} \cite{Ashmore1977}        \\
 H$_{\text{2}}$S    		&  300.0 & (6040)     & -10.79    &   -10.67     &  -0.12      &   -10.43                    &  {\em -10.50} \cite{Bieri1982}          \\
 HClO	            		&  434.4 & (10460)    & -11.96    &   -11.83     &  -0.13      &   -11.3                     &  -11.12       \cite{Colbourne1978}      \\
 HCN		    		&  420.9 & (10060)    & -13.92    &   -13.86     &  -0.06      &   -13.64                    &  {\em -13.61} \cite{Kreile1982}         \\
 NH$_{\text{3}}$    		&  420.9 & (10060)    & -11.45    &   -11.38     &  -0.07      &   -10.92                    &  {\em -10.82} \cite{Baumgaertel1989}    \\
 N$_{\text{2}}$H$_{\text{4}}$	&  420.9 & (10060)    & -10.86    &   -10.78     &  -0.08      &   -10.24                    &  {\em -8.98}  \cite{Vovna1975}          \\
 PH$_{\text{3}}$    		&  300.0 & (6040)     & -10.89    &   -10.79     &  -0.10      &   -10.49                    &  {\em -10.59} \cite{Cowley1982}         \\
\hline \hline
\end{tabular}
(*) quasi-particle value corresponds to Hartree-Fock HOMO level.
\end{table*}

		We then consider the impact of the vertex on the self-energy and on the quasi-particle energy for the HOMO level. This level is defined as the highest occupied orbital in the (post-)$GW$ approximation. 
	This coincides with the HF HOMO level for all molecular systems except CS and N$_{\text{2}}$, for which it corresponds to the HOMO-1 level in the HF approximation. 
	Furthermore, the inclusion of the ladder diagrams only in the polarisation propagator produces the so-called $G_0W_0$ with test charge-test charge interactions (labelled $G_0W_0^{\rm{tc-tc}}$ in the following).
	In terms of the diagrams involved, the latter approximation can be obtained from the more general $G_0W_0\Gamma$ by omitting the exchange diagrams in the self-energy (\textit{e.g.} those in the bottom row in Figs. \ref{Fig:SE_2ndO} - \ref{Fig:SE_infO} above).
	For this level of theory we observe the same level alignment between the quasi-particle HOMO level and HF reference as in the $G_0W_0\Gamma$ calculations, with the sole exception of SiO.
	For this molecule there is a level crossing as in the case of CS and N$_{\text{2}}$.

\begin{figure} 
        	\centering
	        \includegraphics [width=14.0cm] {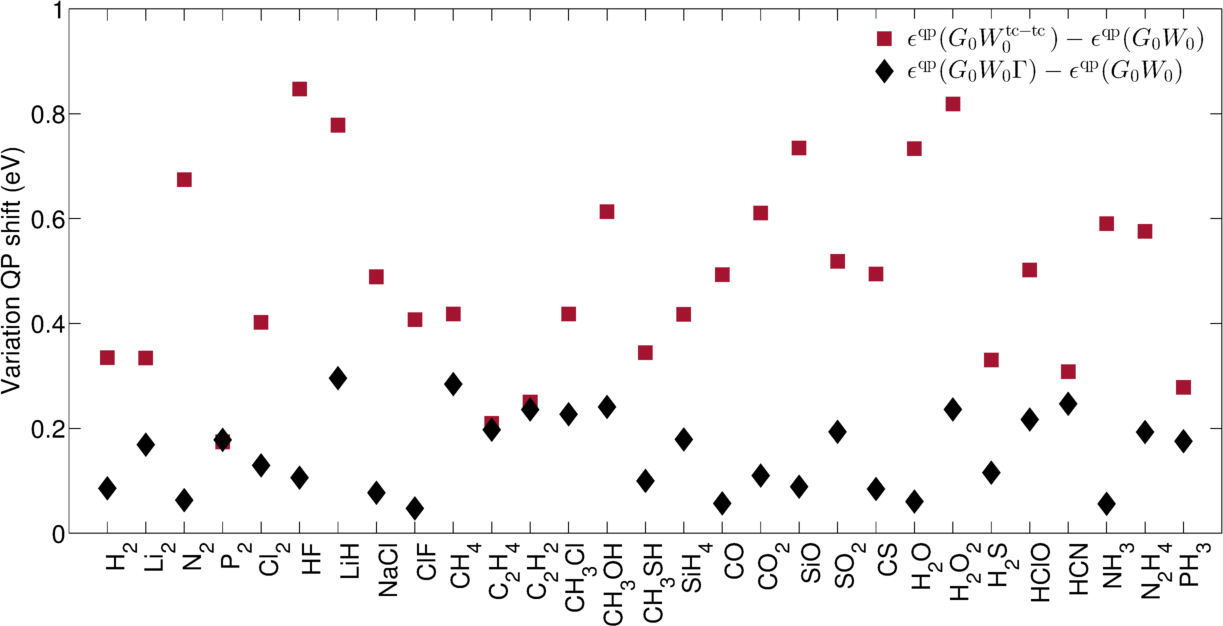}
        	\caption{Variation of the quasi-particle shift for the $G_0W_0^{\rm{tc-tc}}$ (red squares), $G_0W_0\Gamma$ (black diamonds). The reference quasi-particle shifts are taken to be the $G_0W_0$ values.
		Numerical values (in eV) are provided in \autoref{tab:GWGamma}}
	        \label{Fig:varQP}
	\end{figure}

		The inclusion of the ladder diagrams in the polarizability is reasonable from a physical standpoint: it introduces an interaction between the "virtual" particle and hole that are generated as a response of the many-body system to the introduction of a test charge (excitonic effects).
	In turn, ladder diagrams increase the screening in the many-body system, since an interacting electron-hole pair requires less energy to be generated than a non-interacting pair.
	This effect will be present in both the $G_0W_0\Gamma$ and the $G_0W_0^{\rm{tc-tc}}$, and it shows on the resulting self-energy as a shift of its resonances to higher (lower) energies below (above) the Fermi level (conventionally located mid-gap between the HOMO and LUMO levels).
	\autoref{Fig:SE_PH3_CO} contrasts the HOMO self-energy for all approximations considered in two representative systems (PH$_\text{3}$ and CO): in both cases, there is a reduction of the frequency interval devoid of resonances when the polarizability includes ladder diagrams. 
	Consistently the intersection with the bisector line $y=\omega$ occurs in the $G_0W_0^{\rm{tc-tc}}$ approximation at higher (less negative) energies, thus reducing the underscreening error of $G_0W_0$, as shown in \autoref{tab:GWGamma}.
	Not surprisingly, the most sizable changes occur for molecules whose highest occupied orbital (HOMO) has a non-bonding or anti-bonding character, \textit{i.e.} that has a larger overlap with the LUMO and higher unoccupied states.
	For instance: H$_\text{2}$O, H$_\text{2}$O$_\text{2}$, HF, CO$_\text{2}$, NH$_\text{3}$, have quasi-particle $G_0W_0^{\rm{tc-tc}}$ shifts more positive than the $G_0W_0$ values by more than 600 meV as reported in \autoref{Fig:varQP}. 

\begin{table*}
\caption{HOMO quasi-particle energies on Hartree-Fock reference for all levels of theory considered. 
Finite box corrections are included for all calculations as specified in \autoref{sec:Comp}.
All values in eV.
}
\label{tab:GWGamma}
\begin{tabularx}{\textwidth} {@{}l *8{>{\centering\arraybackslash}X}@{}}
\hline \hline
      molecule & \multicolumn{4}{c}{finite basis set} & \multicolumn{3}{c}{extrapolated results}  \\ 
               			&   HF    	&   $G_0W_0$ &  $G_0W_0^{\rm{tc-tc}}$ & $G_0W_0\Gamma$  &  $G_0W_0$ & $G_0W_0^{\rm{tc}}$ & $G_0W_0\Gamma$       \\
                                                             
\hline                                                       
 H$_{\text{2}}$	 		& 	-16.17 &    -16.34      &  -16.00        &  -16.25        & -16.72    &     -16.40       &     -16.52        \\
 Li$_{\text{2}}$    		&     	-4.90  &    -5.26       &  -5.00         &  -5.16         & -5.29     &     -5.05        &      ---          \\
 N$_{\text{2}}$	    		&      	-17.16 &    -16.12      &  -15.45        &  -16.06        & -16.56    &     -15.92       &     -16.39        \\
 P$_{\text{2}}$			&    	-10.61 &    -11.10      &  -10.93        &  -10.92        & -11.31    &      ---         &      ---          \\ 
 Cl$_{\text{2}}$    		&      	-12.06 &    -11.65      &  -11.25        &  -11.52        & -12.08    &     -11.71       &     -11.80        \\
                                                                                                                                         
 HF		    		&      	-17.63 &    -15.83      &  -14.98        &  -15.72        & -16.29    &     -15.43       &     -16.18        \\
 LiH	            		&      	-8.14  &    -8.12       &  -7.39         &  -7.87         & -8.26     &     -7.50        &     -7.94         \\
 NaCl		    		&      	-9.57  &    -9.14       &  -8.65         &  -9.06         & -9.51     &     -9.06        &     -9.32         \\
 ClF				& 	-13.50 &    -13.21      &  -12.78        &  -13.14        & -13.49    &     -13.06       &     -13.33        \\
                                                                                                                                         
 CH$_{\text{4}}$    		&     	-14.80 &    -14.65      &  -14.24        &  -14.37        & -14.95    &     -14.55       &     -14.57        \\
 C$_{\text{2}}$H$_{\text{4}}$	&       -10.23 &    -10.69      &  -10.48        &  -10.49        & -10.91    &     -10.71       &     -10.66        \\
 C$_{\text{2}}$H$_{\text{2}}$	&       -11.09 &    -11.50      &  -11.25        &  -11.26        & -11.73    &     -11.49       &     -11.43        \\
 CH$_{\text{3}}$Cl  		&       -11.82 &    -11.60      &  -11.18        &  -11.37        & -11.90    &     -11.50       &     -11.56        \\
 CH$_{\text{3}}$OH		& 	-12.27 &    -11.30      &  -10.68        &  -11.06        & -11.71    &     -11.10       &     -11.39        \\
 CH$_{\text{3}}$SH		&    	-9.68  &    -9.72       &  -9.37         &  -9.62         & -9.93     &     -9.60        &     -9.76         \\
 SiH$_{\text{4}}$		& 	-13.17 &    -13.12      &  -12.70        &  -12.94        & -13.40    &     -12.98       &     -12.88        \\
                                                                                                                                         
 CO		    		&     	-15.15 &    -14.77      &  -14.28        &  -14.71        & -15.03    &     -14.55       &     -14.89        \\
 CO$_{\text{2}}$    		&      	-14.77 &    -13.94      &  -13.33        &  -13.83        & -14.35    &     -13.75       &     -14.16        \\
 SiO				& 	-12.65 &    -11.72      &  -10.99        &  -11.63        & -12.05    &     -11.30       &     -11.78        \\
 SO$_{\text{2}}$		& 	-13.56 &    -12.83      &  -12.31        &  -12.64        & -13.20    &      ---         &      ---          \\
 CS				& 	-12.78 &    -12.37      &  -11.88        &  -12.29        & -12.63    &     -12.14       &     -12.47        \\
 H$_{\text{2}}$O    		&       -13.85 &    -12.61      &  -11.87        &  -12.55        & -13.10    &     -12.30       &     -12.92        \\
 H$_{\text{2}}$O$_{\text{2}}$	&       -13.16 &    -11.63      &  -10.81        &  -11.39        & -12.12    &     -11.31       &     -11.81        \\
 H$_{\text{2}}$S    		&       -10.45 &    -10.51      &  -10.18        &  -10.39        & -10.79    &     -10.47       &     -10.60        \\
 HClO	            		&       -12.14 &    -11.63      &  -11.12        &  -11.41        & -11.96    &     -11.48       &     -11.66        \\
 HCN		    		&       -13.31 &    -13.65      &  -13.34        &  -13.40        & -13.92    &     -13.64       &     -13.61        \\
 NH$_{\text{3}}$    		&       -11.70 &    -11.10      &  -10.51        &  -11.04        & -11.45    &     -10.86       &     -11.28        \\
 N$_{\text{2}}$H$_{\text{4}}$	&       -11.16 &    -10.47      &  -9.90         &  -10.28        & -10.86    &     -10.31       &     -10.58        \\
 PH$_{\text{3}}$    		&       -10.45 &    -10.64      &  -10.37        &  -10.47        & -10.89    &     -10.62       &     -10.66        \\
 MAE \textit{vs} CCSD(T)	&		&		&		&		&  0.46 $\pm$ 0.16& 0.15$\pm$ 0.15 &	0.24$\pm$ 0.19		\\
 MSD \textit{vs} CCSD(T)	&		&		&		&		&	-0.46	&	0.06	&	-0.24		\\
\hline \hline
\end{tabularx}
\end{table*}

		On the other hand, in the $G_0W_0\Gamma$ approximation (shown in \autoref{Fig:SE_PH3_CO} by dashed-dotted lines) there is another effect at play: the inclusion of exchange diagrams in the self-energy redistributes the spectral weight of its resonances across the frequency range.
	This effect, for the vast majority of systems considered here (the only exceptions being the unsaturated hydrocarbons C$_{\text{2}}$H$_{\text{4}}$ and C$_{\text{2}}$H$_{\text{2}}$), counteracts the upshift of $G_0W_0^{\rm{tc-tc}}$ in comparison with $G_0W_0$ quasi-particle energies.
	In particular the effect is more pronounced for the set of molecules that showed larger quasi-particle shifts from $G_0W_0$ to  $G_0W_0^{\rm{tc-tc}}$ (see \autoref{Fig:varQP}).
	This comparison can be systematically carried out  as shown in \autoref{Fig:varQP} for the whole range of molecules considered: the impact of vertex corrections in the self-energy, \textit{i.e.} in going from the $G_0W_0^{\rm{tc-tc}}$ to the $G_0W_0\Gamma$ approximation, is system dependent.
	In particular, systems with a more ionic character (for instance fluorides and chlorides) or with unsaturated chemical bonds (\textit{e.g.} N$_{\text{2}}$, CO, CO$_{\text{2}}$) will have the electron density rather localised around the more electronegative element or in the bonding region respectively.
	As electrons become more localised correlations among them become pivotal and necessitate the inclusion of the vertex at the self-energy level for an accurate description. 

        Comparison with previous literature is not straightforward. 
	Our observations are in contrast with the well-established behaviour of the local vertex in the $GW \Lambda_{\rm{LDA}}$ approximation, which results in a rigid shift of the quasi-particle energies \cite{Morris2007,Hung2016,Delsole1994}, regardless of the system's details.
        $GW\Gamma$ calculations have also been reported by Shirley and Martin \cite{Shirley1993} for neutral atoms and ions.
	Specifically, it has been reported that in ionized atoms the importance of exchange effects in the polarisability exceeds that of correlation effects in the self-energy, whereas for (neutral) atoms correlation effects in the self-energy are more relevant. With the very diverse systems considered here, we can not substantiate this trend.
 
\begin{figure} 
        	\centering
	        \includegraphics [width=7.5cm] {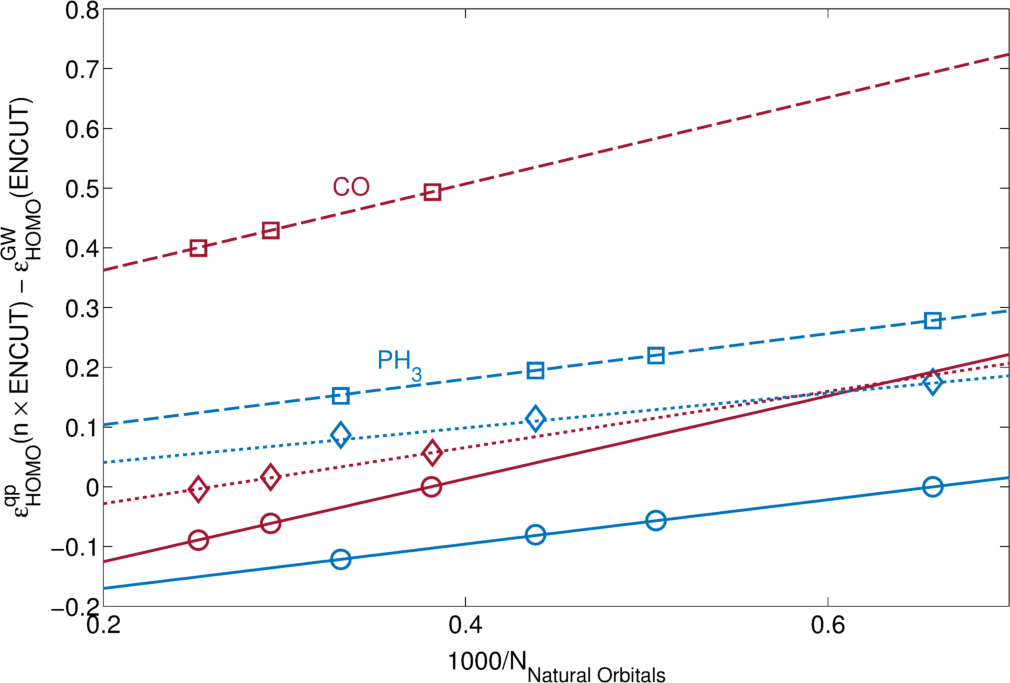}
        	\caption{Quasi-particle HOMO energies for CO (red) and $\textrm{PH}_\textrm{3}$ (blue) as a function of the basis set size.
	The symbols correspond to the quasi-particle energies obtained with the default energy cutoff $E_c$ increased by a factor $n=$1.0, 1.3, 1.5 and, for PH$_\textrm{3}$, 2.0.
	The energy zero has been set to the $G_0W_0$ HOMO energy (with $n=$1.0) for all levels of theory considered: $G_0W_0$ (circles), $G_0W_0^{\rm{tc-tc}}$ (squares) and $G_0W_0\Gamma$ (diamonds). 
	Linear extrapolations are shown as a guide to the eye.}
	        \label{Fig:QP_extrap}
	\end{figure}

		Now we evaluate the basis set corrections for the various levels of theory.
	We notice these values are consistenly smaller for $G_0W_0\Gamma$ than for the $G_0W_0^{\rm{tc-tc}}$ level of theory (see the finite basis quasi-particle levels and the extrapolated counterparts in \autoref{tab:GWGamma}).
	This observation can be rationalised with a similar diagrammatic analysis as carried out for the paramagnetic electron gas \cite{Maggio2016}.
	The argument can be concisely restated as follows: in lowest order there is a cancellation between the bubble and the exchange diagrams in the self-energy (see top and bottom row in \autoref{Fig:SE_2ndO}) because of the sign rule for fermionic loops that assigns a different sign to the two classes of diagrams.
	This cancellation is only partial in second order because there are twice more ways to assign a spin variable to each propagator line in the bubble diagram than there are for the exchange diagram (this is a consequence of the spin conservation at each interaction line).
	From these considerations the linear extrapolation of the quasi-particle energy will have a smaller slope by a factor $\frac{1}{2}$ if the exchange diagrams in the self-energy are included.
	The inclusion of higher order diagrams clearly partially invalidates the previous arguments, since already in third order there are exchange type diagrams that feature the same number of fermionic loops as the bubble diagram (see \autoref{Fig:SE_3rdO} bottom right).
	In general terms, we can however still expect a faster convergence over the number of orbitals for the $G_0W_0\Gamma$ level of theory in comparison with the $G_0W_0^{\rm{tc-tc}}$ case.
	This turns out to be the case for the vast majority of the systems considered here.
	The faster convergence for the $G_0W_0\Gamma$ results can be seen in \autoref{Fig:QP_extrap} for CO and PH$_\textrm{3}$.

		 Finally, we can compare the resulting basis set extrapolated quasi-particle energies with the CCSD(T) results provided in \cite{Bruneval2013}.
	We reiterate that this comparison is somewhat problematic, since basis  set incompleteness errors were not included in the coupled cluster results, and they  are estimated to be over the order of 100 to 150~meV, as already elaborated above.  
	Since CCSD(T) calculations also include exchange diagrams (as all interactions are antisymmetrised) we expect CCSD(T) to have a similar convergence rate as $G_0W_0\Gamma$ and not as $G_0W_0$. 
	Neglecting these corrections, it is clear that the standard $G_0W_0$@HF yields much too large IP's. 
	The $G_0W_0^{\rm{tc-tc}}$ approximation significantly improves on these results, with only few outliers with an error exceeding 300~meV.
	These outliers are  two polar molecules, with much too small IP, namely HF and LiH, as well as CS, CO and N$_\text{2}$ where the IP is still way too large. 
	However, the  $G_0W_0^{\rm{tc-tc}}$  approximation is not a systematic approximation, since it includes vertex correction in the polarizability only  but neglects them in the self-energy. If we keep in mind,
that basis set converged CCSD(T) calculations will result in more negative HOMO energies (about $-100$~meV to $-150$~meV), this could possibly
improve agreement between $G_0W_0\Gamma$ and  CCSD(T) and worsen the  $G_0W_0^{\rm{tc-tc}}$ results.
	
\begin{figure} 
        	\centering
	        \includegraphics [width=14.0cm] {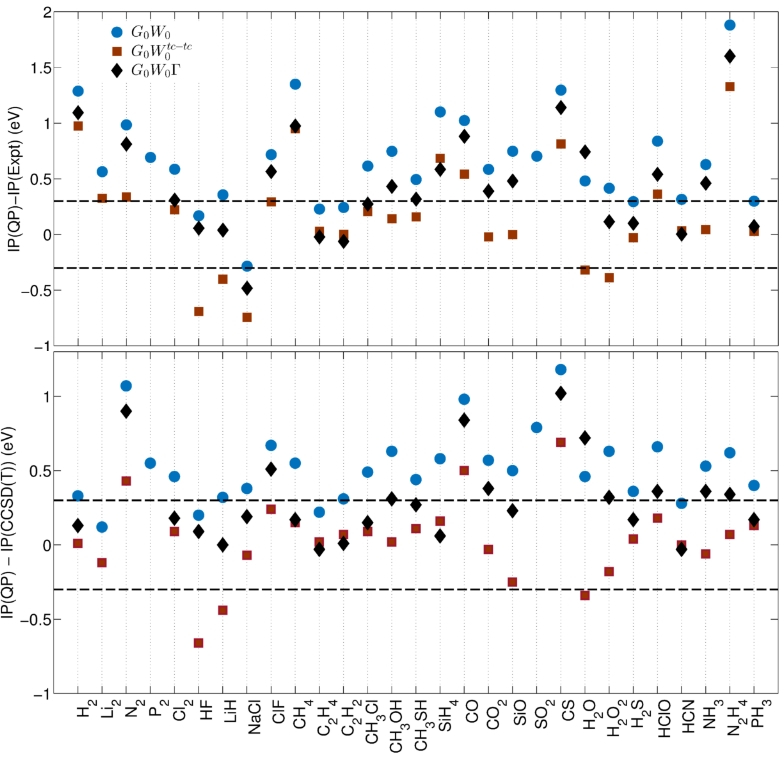}
        	\caption{Deviation from the experimental ionisation potential (top panel) and from the CCSD(T) estimate \cite{Bruneval2013} (bottom panel) for the first ionisation potential ($-\epsilon_{\rm{HOMO}}^{\rm{QP}}$) computed at the $G_0W_0$ (blue circles), $G_0W_0^{\rm{tc-tc}}$ (red squares) and $G_0W_0\Gamma$ (black diamonds) levels of theory.	All computed values include basis set extrapolation corrections. Horizontal lines are centered at $\pm$0.3 eV.}
	        \label{Fig:QPvsCC}
	\end{figure}

		The effect of the vertex in the self-energy is to systematically bring the extrapolated IP again above the CCSD(T) reference values \autoref{Fig:QPvsCC}.
	Generally, the $G_0W_0\Gamma$ IP's lie between the $G_0W_0$ and $G_0W_0^{\rm{tc-tc}}$ values, which is in agreement with the general assumptions in literature.
	Exceptions to this trend are the unsaturated hydrocarbons, SiH$_\text{4}$ and HCN, whose IP's, accidentally, reproduce the CCSD(T) values quite well.
	It is interesting to comment on the biggest outliers in \autoref{Fig:QPvsCC}. These are N$_\text{2}$, CO and CS.
	A previous study \cite{Koval2014} has established the strong dependence on the used  orbitals for carbon monoxide and the nitrogen dimer.
	Indeed, if full self-consistency is achieved (with the update of the initial Hartree-Fock orbitals), the resulting IP's are reduced by an excess of 0.8 eV.
	Unfortunately, similar studies have not been carried out for CS, however we expect a similar behaviour for this system, given the similarities in electronic structures.
	Therefore, for these molecules we expect that an update of the orbitals, or at least of the orbital's energies, more than the inclusion of the vertex will be relevant for an estimate of their IP's.
	
	The mean absolute deviation from the coupled cluster values for the remaining molecules in the set is 0.21 eV for $G_0W_0\Gamma$ and 0.15 eV for $G_0W_0^{\rm{tc-tc}}$, scoring a marked improvement in comparison with the $G_0W_0$ deviation (0.46 eV). 
	We believe that inclusion of basis set corrections in the CCSD(T) results would improve the agreement between $G_0W_0\Gamma$ and CCSD(T) even further, since this would increase the CCSD(T) IP values by the already quoted 100-150 meV moving them closer to the  $G_0W_0\Gamma$. 
	Obviously this correction would worsen the agreement between CCSD(T) and $G_0W_0^{\rm{tc-tc}}$,	which is to be expected, since $G_0W_0^{\rm{tc-tc}}$ is not a well balanced diagrammatic theory.	
	In Ref. \cite{Gallandi2016} range-separated density functionals have been used as a starting point of the $G_0W_0$ calculation, and these scored a similar improvement in quasi-particle energies as the one reported here by applying the vertex. 
	In some sense, this starting point is constructed to reproduce the system's IP, whereas the method we employ is completely \textit{ab initio} and, as such, does not require any previous information on the system's electronic structure.

	Another well established class of \textit{ab initio} approaches is the algebraic diagrammatic construction (ADC). It offers a general framework to generate the self-energy diagrams at any order in perturbation theory.
	The two-particle one-hole Tamm-Dancoff approximation (2ph-TDA) is the simplest ADC method and it includes bubble and particle-hole ladder diagrams in its resummation.
	It is quite striking that, albeit this approximation is diagrammatically similar to the $GW\Gamma$ method, its performance for small molecules is unsatisfactory with a mean absolute error (\textit{vs} CCSD(T)) that exceeds 0.6 eV \cite{Corzo2015}.
	The inclusion of all self-energy diagrams up to third order, as it is performed in the ADC(3) approximation, largely improves the IP's values.
	It is important to point out that in the 2ph-TDA approach the 2p-2h virtual excitations are not included, whereas these effects are implicitly accounted for by the coupling amplitudes in the full ADC(3) \cite{VonNiessen1984}. 
	These excitations are responsible for introducing correlation effects in the HF ground state \cite{Ring1980} (which is obviously uncorrelated) and are explicitly built in the $GW$ and $GW\Gamma$ approximation thanks to the inclusion of the so-called resonant-antiresonant coupling.
	The comparison between our results and those obtained with the electron propagator methods cited above allows us to rationalise the poor performance of the 2ph-TDA.
	Specifically, the coupling between resonant and antiresonant virtual transistions is a key feature of $GW$ (and $GW\Gamma$) that is implicitly accounted for in ADC(3).
	This coupling seems to be responsible for a dramatic improvement over the quasi-particle energies obtained at the 2ph-TDA level.
	On the other hand, the inclusion of other higher order particle-particle ladders in the ADC(3) method appear to have an overall modest impact (with the exceptions discussed above that are poorly described by the HF reference), since its performance on small molecules is comparable to that of the $GW\Gamma$ method. 
	
	The comparison with experimental data (\autoref{Fig:QPvsCC}, bottom panel) proceeds along similar lines as above, with the mean absolute error increasing slightly owing to the impact of adiabatic effects on the measurements for a number of systems considered (these are reported in \autoref{tab:G0W0} as non-italicised entries).

\section{Conclusions}

	The main topic of this work is the inclusion of  exchange diagrams in both the polarization propagator as well as the self-energy using Hedin's equations. 
	Commonly this is referred to as the $GW\Gamma$ approximation. 
	In Hedin's equations the interaction kernel is strictly given by the functional derivative of the self-energy employed in the preceding step with respect to the Green's function. 
	In the present work, we adopt the Hartree-Fock starting point. 
	Then  from the derivative of the Hartree-term $V_\mathrm{H}$ with respect to the Green's function, one obtains the standard RPA bubble diagrams, and from the derivative of the exchange potential $\Sigma_\mathrm{x}=V_\mathrm{x}$ one obtains the ladder diagrams. 
	Hence, applying the common $GW$ approximation--- that includes bubbles only but neglects ladder diagrams ---is from the outset a flawed prescription for the Hartree-Fock starting point. 
	If it were successful, this could only be related to  some sort of error cancellation.

	An important aspect of the present work is whether such a cancellation exists. 
	To finally and concisely answer this question, we have performed quasi-particle calculations on top of the Hartree-Fock approximation for a set of 29 small molecules. 
	As already alluded to above, we have included the ladder diagrams related to the derivative of the exchange potential, in both the polarization propagator and the self-energy. 
	The applied equations have been derived from Hedin's equations and should be fairly easy to implement in any time-dependent Hartree-Fock, Casida or BSE code. 
	We have also argued that, although the final equations (\ref{eq:SigmaR}) essentially only rephrase the equation of motion for the Green's function,  they have not yet been used in quantum chemistry or computational solid state physics.
	This is somewhat astounding, since the implementation  is, in principle, straightforward. 
	All one needs to know are the two-particle eigenstates of the TD-HF equation. 

	Now let us turn to our results. 
	The first, already often published, observation is that the $G_0W_0$ approximation is not satisfactory when performed on top of HF.
	It systematically overestimates the IP's. This should come as no surprise.
	In the $G_0W_0$ approximation, a non-physical polarizability is used, that is calculated without particle-hole ladder diagrams related to the variation of the exchange. 
	Inclusion of the ladder diagrams in the polarization propagator (which is equivalent to using the polarizability from TD-HF) largely rectifies this issue and massively improves agreement with the CCSD(T) reference values.
	The mean absolute error decreases from about 400~meV in $G_0W_0$ to 150~meV in this $G_0W_0^{\mathrm{tc-tc}}$ approximation.

	Including the exchange diagrams in the polarizability only is, however, not an entirely well balanced approximation instead, one should include the exchange diagrams also in the self-energy ($G_0W_0\Gamma$ approximation). 
	The $G_0W_0\Gamma$ values are always in between the standard $G_0W_0$ and the $G_0W_0^{\mathrm{tc-tc}}$ results, as discussed in more detail in the next paragraph. 
        With a mean absolute deviation of 210~meV compared to the CCSD(T) reference results, $G_0W_0\Gamma$ results are on average slightly worse than the  $G_0W_0^{\mathrm{tc-tc}}$ (MAE 150~meV).
	This is certainly somewhat unsatisfactory and requires further studies and, in particular, better reference values. 
	With an error of roughly 100~meV for cc-pVQZ basis sets, CCSD(T) values without basis set extrapolation are not sufficiently accurate to allow for an unambigious comparison.
	In passing, we also emphasize a side result of our study, namely that $G_0W_0$ converges differently with respect to basis set size than  $G_0W_0\Gamma$.
	This is a result of the inclusion of the exchange diagrams in the self-energy as shown in  Fig. \ref{Fig:QP_extrap}.
	This also means that comparisons at finite basis sets (without basis set corrections) need to be done with caution, since one could observe fortuitous agreement at small basis sets that might not hold up at improved basis sets. 

	A key result of this study is that changes from $G_0W_0^{\mathrm{tc-tc}}$ to $G_0W_0\Gamma$ are very system-dependent: in some cases, $G_0W_0\Gamma$ recovers the $G_0W_0$ values, in other cases the final values are closer to $G_0W_0^{\mathrm{tc-tc}}$. 
	As an example, one can consider the case of ammonia and phosphine: they have a comparable geometry and valence electron configuration, with the obvious difference that phosphorus is one period below nitrogen in the periodic table.
	In one case (PH$_{\text{3}}$) the introduction of the vertex in the self-energy has a marginal impact on the IP, whereas for NH$_{\text{3}}$ the higher electronegativity of the central atom, as discussed in the previous section, leads to increased electron correlations and to a more pronounced difference between  $G_0W_0^{\mathrm{tc-tc}}$ and $G_0W_0\Gamma$.
	Albeit these observations are in line with "chemical intuition", they are difficult to generalise, let alone if quantitative estimates are required. 
	We conclude that there is simply no shortcut to $GW\Gamma$ or general error cancellation; one needs to include the vertex consistently.
	All these results are in agreement with previous work on solids, where on average the vertex in the self-energy also compensated the effect of the vertex in the polarization propagator, and as for molecules, the results varied significantly between different solids and even different orbitals \cite{Gruneis2014a}.

	Finally, let us comment on the future perspectives: to us it is clear that the effect of self-consistency on the Green's function needs to be studied. 
	Unfortunately, this is not a simple matter. 
	If one updates the self-energy and the Green's function, additional terms come into play in the interaction kernel $I$, and ultimately one is then forced to make compromises between computational efficiency and accuracy \cite{Kutepov2016}. 
	We plan to report a combination of self-consistency and vertex corrections in our forthcoming study.

	In summary, we have shown that $G_0W_0\Gamma$@HF improves upon  $G_0W_0$@HF. 
	This implies that the vertex in Hedin's equations is important and can not be neglected. 
	Given the simplicity of the present approach, we are confident that it will draw interest in the computational quantum chemistry community, as it constitutes an important fundamental step towards accurate quasi-particle calculations.

\section*{Acknowledgements}
Funding by the Austrian Science Fund (FWF) within the Spezialforschungs-
bereich F41 (SFB ViCoM) is gratefully acknowledged.
Computations were predominantly performed on the Vienna Scientific Cluster VSC3.

\bibliography{Books,BSE,CoupledCluster,ElectronCorrelations,GW,GeneralTheory,RPA}

\end{document}